# Is there something I'm missing?

Topic modeling in eDiscovery

*Herbert L. Roitblat, Ph.D.*

*Mimecast*

# Is there something I'm missing?

Topic Modeling in eDiscovery

Herbert L. Roitblat, Ph.D.

Mimecast

## Abstract

In legal eDiscovery, the parties are required to search through their electronically stored information to find documents that are relevant to a specific case. Negotiations over the scope of these searches are often based on a fear that something will be missed.  This paper continues an argument that discovery should be based on identifying the facts of a case. If a search process is less than complete (if it has Recall less than 100%), it may still be complete in presenting all of the relevant available topics. In this study, Latent Dirichlet Allocation was used to identify 100 topics from all of the known relevant documents.  The documents were then categorized to about 80% Recall (i.e., 80% of the relevant documents were found by the categorizer, designated the hit set and 20% were missed, designated the missed set).  Despite the fact that less than all of the relevant documents were identified by the categorizer, the documents that were identified contained all of the topics derived from the full set of documents.  This same pattern held whether the categorizer was a naïve Bayes categorizer trained on a random selection of documents or a Support Vector Machine trained with Continuous Active Learning (which focuses evaluation on the most-likely-to-be-relevant documents).  No topics were identified in either categorizer's missed set that were not already seen in the hit set.  Not only is a computer-assisted search process reasonable (as required by the Federal Rules of Civil Procedure), it is also complete when measured by topics.

## Introduction

The parties in a lawsuit have an obligation to perform a reasonable search for relevant documents, and this search process can be highly burdensome and expensive.  The parties in many cases try to negotiate methods and criteria and these negotiations are often disputatious, often because one party or the other is concerned that something important might be missed.  I argued in an earlier paper (Roitblat, 2019) that fear of missing out contributes to the high cost of eDiscovery.

Arguably, the most important contribution of that earlier paper is the idea of viewing the topics in a collection of documents as something that is separable from the documents in the collection.  The general idea is that a document contains one or more topics.  The fact that the same topic is mentioned in multiple documents could mean that a search process has found all of the available relevant topics even it has missed some of the relevant documents. I argued in that paper, that missing documents do not necessarily imply missing important topics. Now I have had an opportunity to evaluate some actual documents. What I found in this new study is that no topics were found in among the missed documents that were not already found in the identified relevant ones.

The general approach of this study is to
1. Identify the topics available in a set of known relevant documents
2. Do machine learning to identify the relevant documents that can be discovered
3. Compare the topics present in the relevant documents with the documents present in those that were not identified by the machine learning.
4. Determine whether any new topics that were identified in the missed set that were not already present in the set identified as relevant by the machine learning.

The basic idea behind topic modeling is that documents consist of words that are derived from some mixture of topics. The goal of eDiscovery, I argue, is to get the information contained in a collection of documents not to get the documents themselves. The documents are just a means to reach the information.

Statistically, we can say that for any collection of documents, there is an underlying, statisticians say "latent," set of topics. Each topic is associated with a distribution of words. That is some words occur more frequently than others with a given topic, and some words occur more frequently with some topics than with others. From a statistical point of view, each document is generated by first picking one of these topics and then selecting words based on this topic's distribution. Any given document may reflect one or more topics.

There are statistical techniques that can be used to work backwards from the words in documents to the topics that were most likely to have generated them. One of these techniques is called [Latent Dirichlet Allocation (](Blei, Ng, & Jordan, 2003). This is one of several standard statistical or machine learning methods for identifying the topics involved in a set of documents. It computes the topics that were most likely to generate each document, based on the words contained in that document.

Latent Dirichlet Allocation, for example, might identify one topic that is legal-related and another that is sports-related. The word "court" might appear in both of them. But "judge" or "tennis" would be much more likely in one topic than in the other.

As in clustering, the topics that Latent Dirichlet Allocation identifies may not align strongly with the topics that a human might suggest, but they do cover the range of topics in the set of documents. In this study, I extracted the set of topics from all of the known relevant documents. I then trained a machine learning categorizer to identify relevant documents in order to work backwards and determine which of the topics was covered by the documents in the identified set (the hit set) and which were covered by the documents that were missed (the missed set).

By extracting the topics from the full set of relevant documents, it is possible to identify the topics that are contained in the set that have been classified as relevant and those in the relevant documents that have not been categorized as relevant. By hypothesis, there should be no or very few topics in the non-identified set that are not also in the identified relevant set.

## The data

The documents for this study consist of a set of micro-aggression communications collected by Luke Breitfeller and colleagues (2019). Microaggressions are statements or actions, that express a prejudiced attitude toward a member of a marginalized minority. These statements can have powerful and long-lasting effects on the people to whom they are directed. I selected these data as part of a project to proactively identify unintentional use of harmful language so that people could identify potential aggressions and alter their language before it can cause discomfort.

Breitfeller and colleagues mounted a substantial effort to collect and identify microaggressive statements from social media. They started with self-reported incidents of microaggressions and then developed methods of annotating further examples. The data consist of social media posts, rather than of more typical emails, so they may be a bit shorter than an average email. See Breitfeller's paper for details. Nevertheless, because they are categorized, they provide a good set of relevant and nonrelevant documents. Because they are brief, they limit the number of topics that can be included in each one and, therefore, provide a more conservative estimate than emails of the number of topics contained in an identified set of relevant documents.

A few example microaggressive statements:
- At least I don't sit on my ass all day collecting welfare, I EARN my money.
- Before we met when I first saw your name I thought you were going to be one of those quiet Asians-you know? And my mom was like 'well maybe she'll just spend all her time in the library so you'll have the room to yourself! I'm so glad you're not like that!
- Did you get this job because you're pregnant?
- Do your parents make sacrifices so that you can go to our school?

I selected all 1,676 microaggressive statements from their collection and a haphazardly selected set of 2,000 other texts from their collection. For this paper I am interested in analyzing the topics of the relevant texts; I'm not so interested in the ability to distinguish relevant from nonrelevant documents. Nonetheless, the difference between microaggressions and other statements can be quite subtle, so this task is challenging for any text classifier.

## Identifying topics

Several methods (e.g., latent semantic analysis, e.g., Deerwester, 1990, probabilistic latent semantic analysis, Hofman, 1999) are available for inferring the topics associated with each document. For this study, I chose, as mentioned, latent Dirichlet allocation (Blei, Ng, & Jordan, 2003). These methods require the researcher to specify the number of expected topics. For this analysis, I chose 100 topics. With 1,676 short documents to work with, specifying a hundred topics seems like a very large number, with an average of about 16.8 documents per topic. The actual number of documents assigned to each topic, however, varied widely (See Figure 1). Some topics encompassed many documents, but most topics included only a few. I will describe this pattern in more detail shortly. The conclusions we can draw with this large number of topics would also apply to any lower number. Larger numbers of topics would imply fewer average documents per cluster. For simplicity, we focused on just one topic per document, the one that was most associated with that document. If the analysis allowed more topics per document, then it would have been even more likely to find all topic among the documents identified by the categorizer.

All of the 1,676 microaggressive statements were analyzed using Latent Dirichlet Analysis with 100 latent topics.

By way of example, the 15 most closely associated words for the first four Microaggression topics are shown in Table 1.

*Table 1. The top words for the first for topics identified using Latent Dirichlet Analysis*

| Topic 0 | tell, mixed, person, ask, race, knew, woman, mom, want, feel, white, oh, black, know, excellent |
| Topic 1 | student, female, street, deserve, sorry, male, sir, miss, people, know, american, black, african, difference, theres |
| Topic 2 | exotic, walk, meaning, harder, means, gets, place, muslim, want, country, students, woman, thought, native, american |
| Topic 3 | life, know, english, woman, sisters, friend, right, white, really, older, teacher, language, oh, youre, born |

For each microagressive statement, the topic with highest association score was selected. For example, the first microaggressive statement quoted above ("At least I don't …") was most strongly associated with Topic 16 (0.7525), and weakly associated (0.0025) with the 99 other topics. Each number in the following list indicates the strength of the association between the microaggressive statement and the corresponding topic. The first topic is Topic 0.

0.0025 0.0025 0.0025 0.0025 0.0025 0.0025 0.0025 0.0025 0.0025 0.0025
 0.0025 0.0025 0.0025 0.0025 0.0025 0.0025 **0.7525** 0.0025 0.0025 0.0025
 0.0025 0.0025 0.0025 0.0025 0.0025 0.0025 0.0025 0.0025 0.0025 0.0025
 0.0025 0.0025 0.0025 0.0025 0.0025 0.0025 0.0025 0.0025 0.0025 0.0025
 0.0025 0.0025 0.0025 0.0025 0.0025 0.0025 0.0025 0.0025 0.0025 0.0025
 0.0025 0.0025 0.0025 0.0025 0.0025 0.0025 0.0025 0.0025 0.0025 0.0025
 0.0025 0.0025 0.0025 0.0025 0.0025 0.0025 0.0025 0.0025 0.0025 0.0025
 0.0025 0.0025 0.0025 0.0025 0.0025 0.0025 0.0025 0.0025 0.0025 0.0025
 0.0025 0.0025 0.0025 0.0025 0.0025 0.0025 0.0025 0.0025 0.0025 0.0025
 0.0025 0.0025 0.0025 0.0025 0.0025 0.0025 0.0025 0.0025 0.0025 0.0025

The number of aggressive statements associated with each topic are shown in Figure 1.

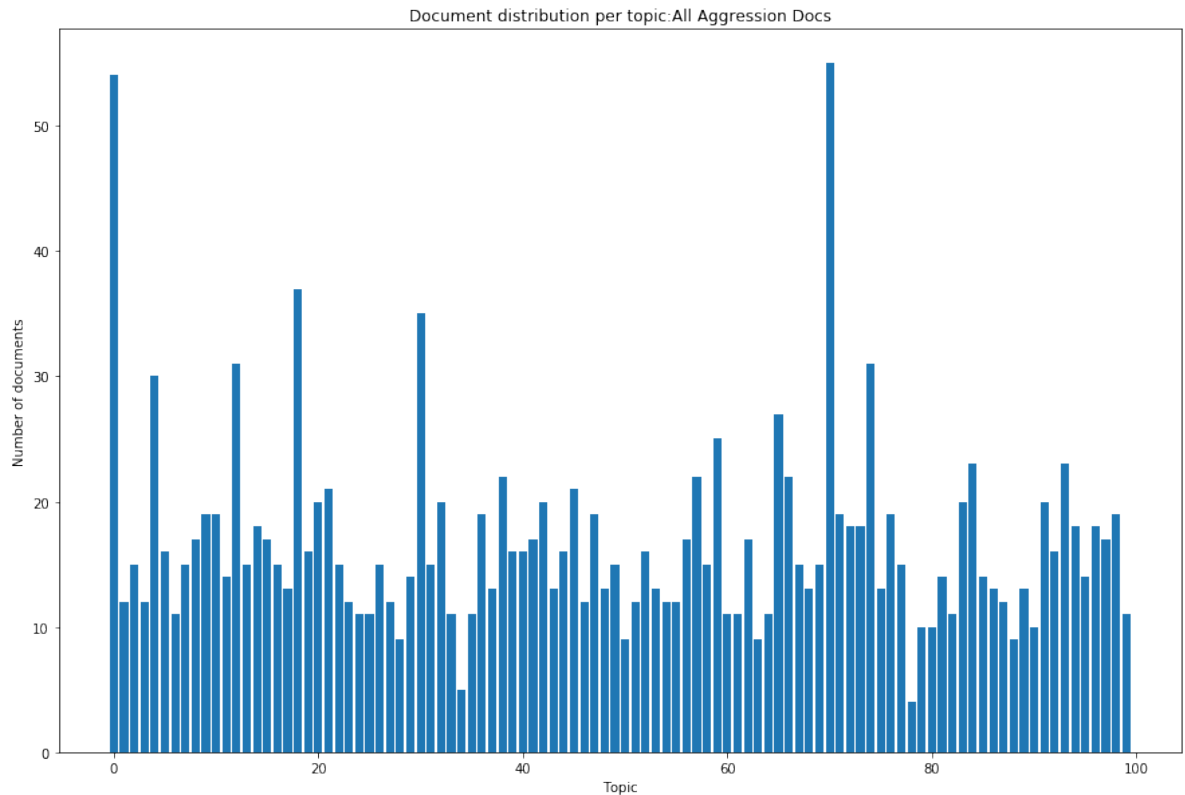

*Figure 1. The distribution of microaggression statements per each of 100 topics for all documents.*

These are the topics associated with the full collection of aggressive statements. Only the strongest topic association for each document is shown.

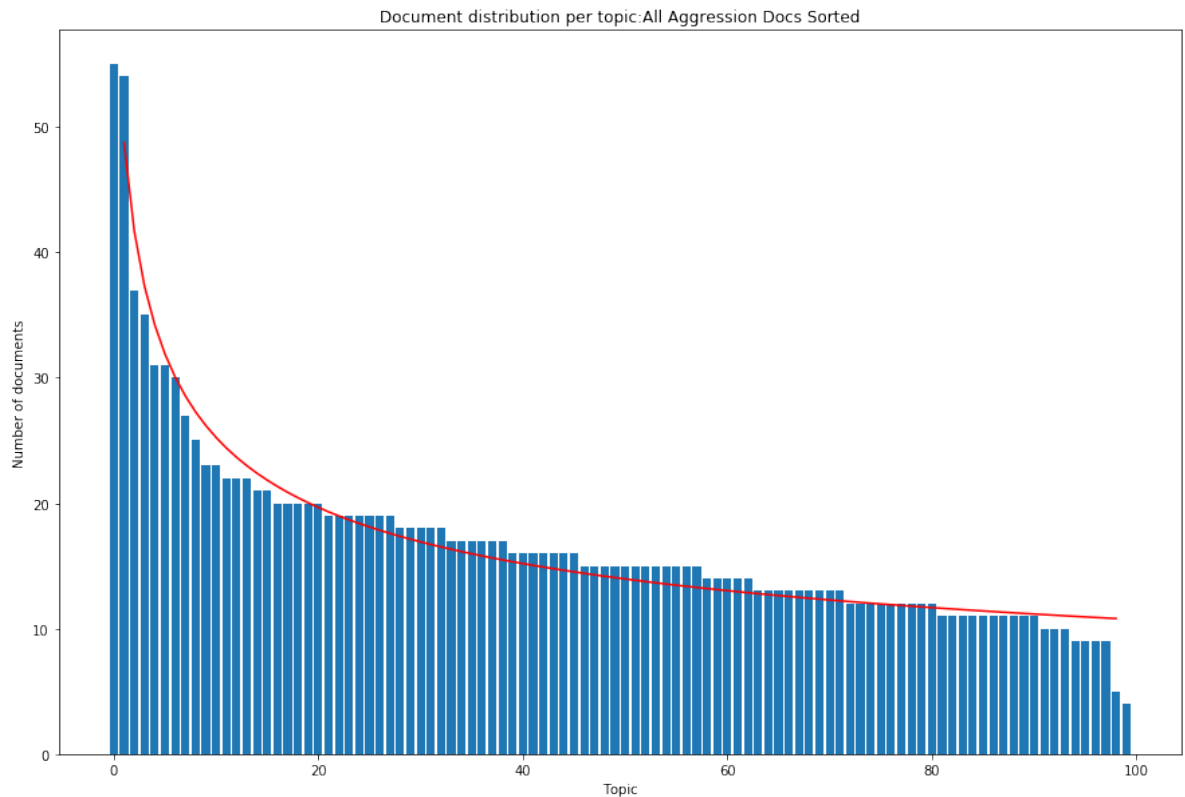

*Figure 2. The number of documents for each topic sorted in descending order. The red line shows the expected number of documents under the predicted power-law distribution.*

Figure 2 shows the categories sorted in descending order of the number of documents. As predicted in the earlier paper, the number of documents corresponds to a power-law distribution. Some topics have many more documents than others

## Machine learning

It is common in eDiscovery to use some form of supervised machine learning to identify the responsive or relevant documents in a collection. In this section I used a version of naïve Bayes classifier. This classifier works by identifying the words that are characteristic of each class of documents, here, the relevant (microaggressive) and the non-relevant (non-aggressive) classes. Training consists of extracting the words from example documents of each category and tracking the frequency with which each word occurs in each class. The categorizer, like all categorizers, depends on having a set of annotated example documents to work from. Annotated, in this context, means that each document in the training set is labeled for the class to which it belongs.

There are several methods for getting these annotations. In this case, we leveraged the work done by Breitfeller and colleagues. The training examples were provided to machine using a standard approach of randomly assigning documents from both classes to either the training set or to a testing set. A randomly selected 1,257 microaggression and 1,500 non-relevant statements were used for training and the rest were used for evaluating the learning process. These remaining documents are called the holdout set. No topic information was used during this phase of the study.

Initially, the quality of the naïve Bayesian classifier was assessed against the hold-out set that was not used during training. After training on these 2,757 training documents, the system yielded 0.75% Recall and 85% Precision against the holdout set.

The next phase of the study examined all of the statements that had been used as positive training examples and those that were identified in the holdout set as being relevant. These are the documents that would be produced if the goal was to identify those documents that represent microaggressions. A total of 1,358 relevant documents emerged from the training and holdout set (the hits). A total of 318 documents were misclassified as non-relevant (the misses). By this measure the combination of training and testing documents yielded a Recall level of 81.0%. That is, the categorizer was able to identify 81% of the relevant documents in this collection.

I then examined which topics were represented by the documents in these two sets. Figure 3 shows the distribution of topics for the hits. Obviously, some topics were associated with more topics than some others.

All 100 topics were represented among the hits (counting only the presence of the strongest topic). As a result, the hits identified all of the available topics even though the machine learning process failed to identify 381 documents. No topics were missed among the hits, so no topics were available to be discovered only in the missed set. The hit set was sufficient to cover all 100 topics.

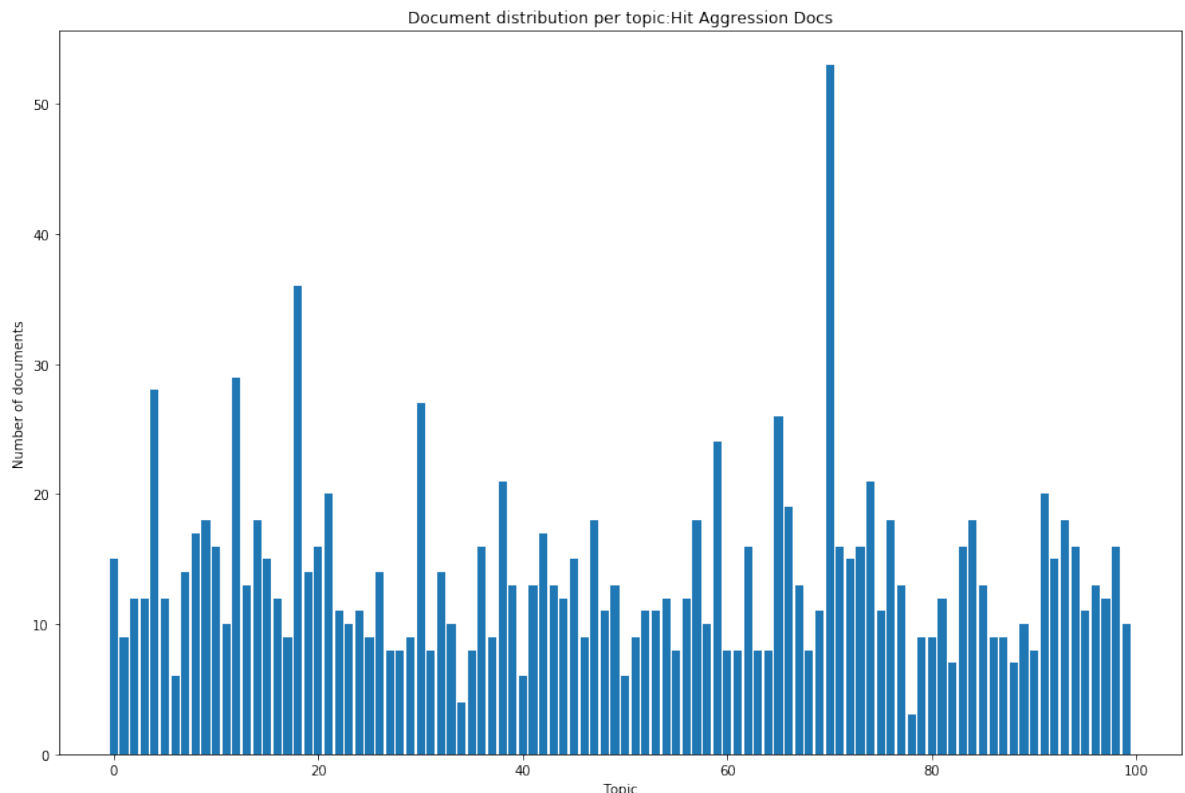

*Figure 3. The number of documents holding each topic from among those that were misclassified as relevant by the Naive Bayes Classifier. Only the documents in the hit set are shown.*

Figure 4 shows that the small set of 381 missed documents includes 93 of the 100 topics. Therefore, there were few topics that would have been missed, even if the search had produced such a small number of hits. Because all of these topics were found among the hit set, there would be nothing to be gained (measured in terms of additional topics) from attempting to search for the documents that remained after the Naïve Bayes categorizer did its job to 81% Recall.

An argument could be made that the topic completeness of the hit set derives from the random sampling process used to present documents to be categorized. Legal discovery does not always use random selection to provide a training set. For example, continuous active learning (Cormack and Grossman, 2014) presents for labeling the documents that it predicts are most likely to be relevant and these are surely not in random order. To evaluate that argument, I repeated the same observations using a simulation of continuous active learning (Cormack and Grossman, 2014). Instead of a Naïve Bayes classifier, the next set of observations used a Support Vector Machine. Instead of processing a random sample of training documents, the next observation started with a small, randomly chosen set of documents, but then focused on those documents that the machine learning system designated as most likely to be responsive, repeating this pattern over several batches until the system achieved a Recall level (81.3%) comparable to that achieved using the Naïve Bayes system (81.0%). Unlike the ordinary use of continuous active learning in eDiscovery, the review was halted when Recall on the entire collection (not just those documents already observed) reached this criterion. The goal was to test the effect of non-random selection of training documents on the topic coverage.

Even though the documents chosen for training in this simulation were not selected randomly, these results also show that all 100 topics were identified among the documents designated by the machine learning system as relevant. No topics were left to be discovered among the missed documents.

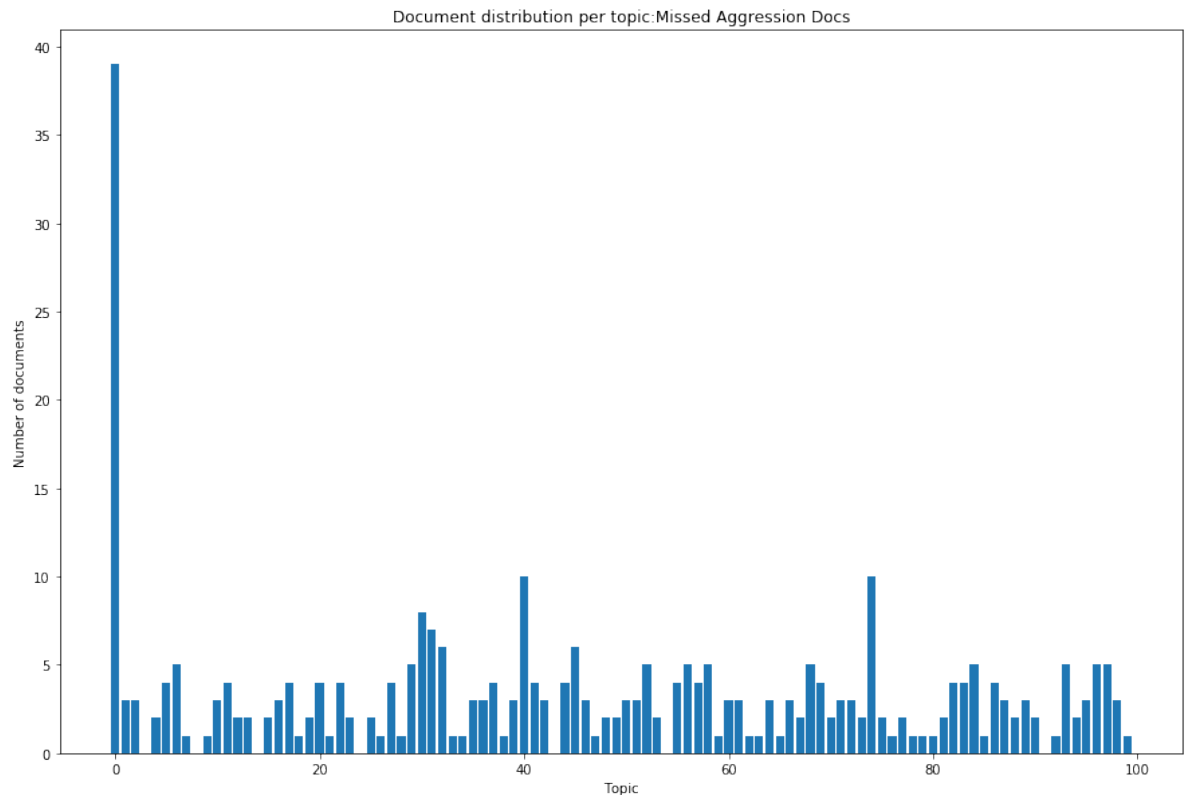

*Figure 4. The number of documents holding each topic from among those that were misclassified as non-relevant by the Naive Bayes Classifier.*

The documents in the missed set of documents contained 92 out of 100 topics. There is, thus, no evidence that non-random document selection has any significant effect on topic coverage among the identified relevant documents.

## Discussion

Predictive coding or machine learning categorization is not guaranteed to find all of the relevant documents in a selection. Instead the goal is to find a reasonable number of these documents after a reasonable effort. There is no hard and fast rule about what reasonable means in this context, but 80% Recall is in the range of many reported projects using machine learning. There is nothing special about 80%, it is just a number in the neighborhood of common experience. What makes a certain Recall reasonable will depend on the precise circumstances of the exercise, the complexity of the distinction between relevant and non-relevant documents, and many other factors.

What this study does show, however, is that 80% Recall is reasonable in another sense. It appears that this level of Recall is sufficient to find all of the available topics in a set of relevant documents. Lower levels of Recall may also be sufficient, but the present results suggest that higher levels may not be necessary to identify all of the relevant topics in a set of relevant documents.

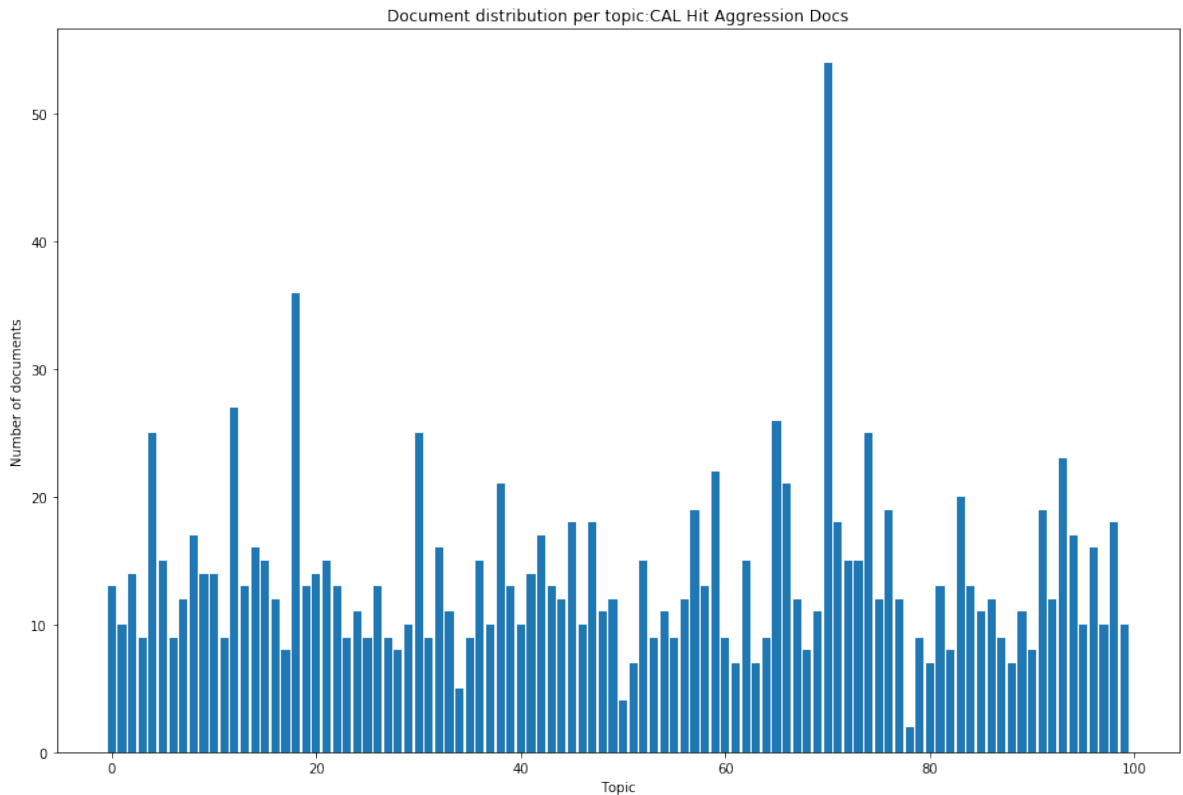

*Figure 5. The number of documents holding each topic among the identified set resulting from continuous active learning and a support vector machine.*

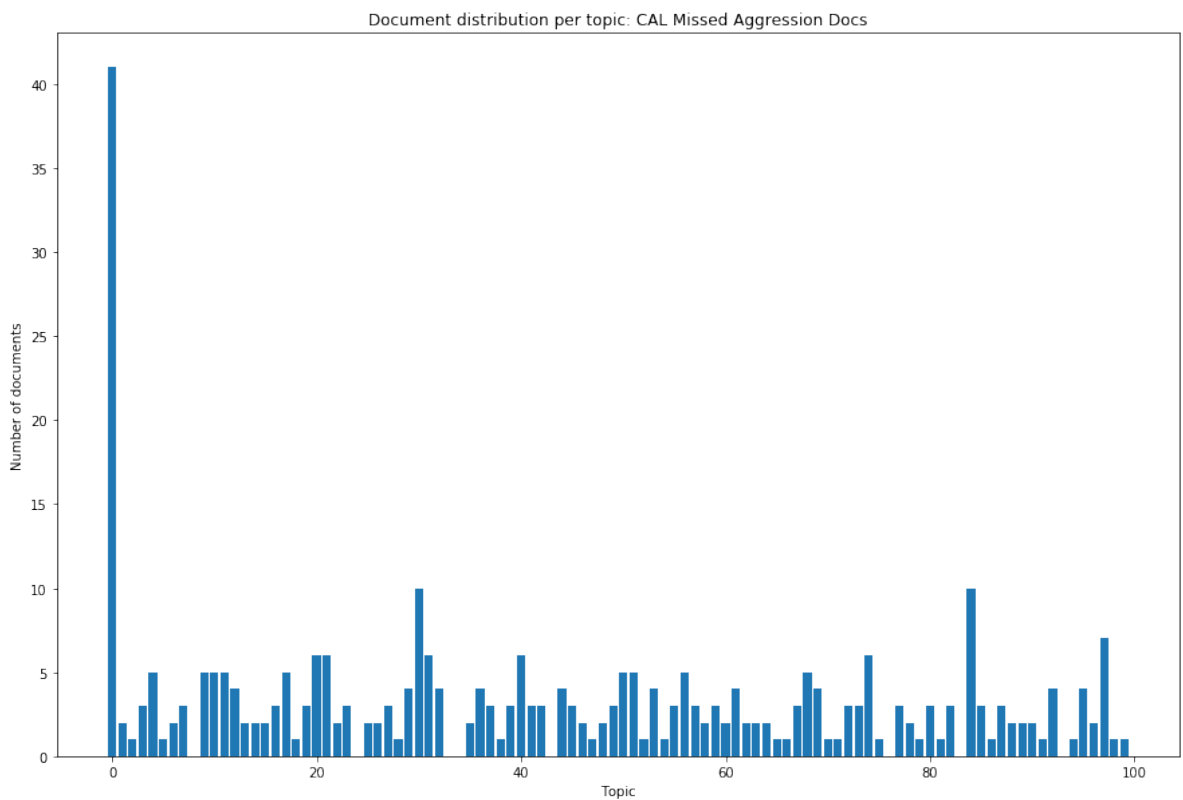

*Figure 6. The number of documents holding each topic from among those that were misclassified as non-relevant by the Support Vector Machine*

Consistent with my earlier paper, we find no evidence that there is significant value in continuing to search for new information once a moderate level of Recall has been achieved (say around 80%). The previous paper was based solely on mathematical modeling of the probabilities and confidence levels associated with a variety of processes. This study used real data to verify some of the properties of that previous modeling.

The same results were found using a Naïve Bayes Classifier trained on a random selection of documents a simulation of a currently popular eDiscovery method called continuous active learning. There are two main differences between Continuous Active Learning and the Naïve Bayes model described earlier. The continuous active learning system selects documents for labeling according to their predicted likelihood of being responsive and then trains machine learning method on the basis of those labeled documents using a support vector machine. Although the same labeling selection pattern could be used with any machine learning method, the current Naïve Bayes implementation used a random selection scheme. In both cases, all of the available topics were discovered among the documents predicted to be responsive. The ordered selection of documents for labeling (and including in the training set) had no discernible effect on the measure of interest, whether the documents predicted to be responsive were sufficient to identify the available topics.

It is striking that with even as few as 314 missed relevant communications, were enough to capture 92% of the identified topics in the Continuous Active Learning experiment. Of the 318 missed documents in the Naïve Bayes Classifier part, still allowed identification of 93 of 100 topics. These results lend confidence to the idea that 100 topics was a reasonable number for the full set. Most of them could be found with even a very small set of documents.

These results are in spite of several factors that might have prevented topics from being discovered, including that the documents used in this study were relatively brief, that there were so few of them. eDiscovery exercises tend to produce rather larger sets of relevant documents, typically numbering in the tens or even hundreds of thousands.

Every study has its limitations and it might be worthwhile to consider some of those here. One of these limitations might be the selected number of topics. Despite the suggestion that 100 topics was enough, one could always claim that what we really needed was more topics.

Although the selected number of topics was large relative to the number of relevant documents, it is possible that we could find some large number of topics that would not be completely covered by the hit set. In this study, however, we had the advantage of knowing all of the relevant documents in the collection and the topics were derived from this set. So more topics might have subdivided the documents further, but it is unclear that they would have had much, if any, of an effect on whether a topic was covered by the hit set versus the missed set. Further research could be applied to answer this question.

In any case, it is not clear how many topics would be appropriate for an eDiscovery exercise. Of the thousands of documents identified as relevant and produced in a typical eDiscovery exercise, maybe as few as a dozen of these are ultimately used in court. When the lawyers have categorized relevant documents, most of the cases I have seen have limited themselves to only a few tens of category tags. The most topic tags I have seen in an eDiscovery case was designed to be about 300 categories, but only a few of these were actually used reliably during the document review. Because the topics in the present study are derived from the total set of relevant documents, adding more topics may serve only to subdivide the documents further into overlapping topics without adding any new value.

The current study was conducted on a relatively small set of documents, each of which tended to be relatively brief.  This small size made this study practical, but it also required some other adaptations.  Actual eDiscovery collections can contain many thousands of relevant documents, which would, in principle, make it easier to find examples of each topic.

The foundational conclusion of these experiments is that a reasonably conducted search of a document collection is very likely to yield examples of each of the identifiable topics of a case.  Documents may be missed, but based on mathematical modeling and the present observations, it is unlikely that topics will be missed.

The final part of the present study demonstrates that this pattern of results does not depend strongly on random sampling of the training documents. A nonrandom training regimen produces essentially the same results as a random one. Based on some additional modeling, to be described in a separate paper, even substantial departures from random selection do not substantially affect the overall pattern of results, this study confirmed that one common way of sequencing the documents used to train the model did not have any obvious effect on the number of potential causes identified.

In retrospect, it is perfectly reasonable to expect similar topic coverage for different methods of training document ordering.  Two methods that achieve 80% overlap with the set of relevant documents must have substantial overlap with one another.  Both the naïve Bayes classifier and the continuous active learning systems correlated to the same level (about 80% Recall) with the set of all relevant documents, so they would be expected to overlap in about 64% or more of those documents, and that would be enough to cover all of the topics.

The focus on topics as the appropriate unit of discovery may be seen as a radical departure from ordinary practice.  However, it addresses directly the concept of whether there is likely to be new and useful information in documents that are predicted to exist (because Recall is estimated to be less than 100%) but not yet identified after a reasonable search.  The analyses presented here and in the earlier paper suggest that there is little new information to be gained from the disproportionate effort needed to search for additional relevant documents among those that were missed by the initial discovery search.

Nothing here says that there cannot be some new topic or some fact that was not already identified in an actual eDiscovery exercise, but it does say that such new topics are unlikely. These studies have not proved there are no unicorns to be found in the missed set, but finding one will be unlikely, closer to wishful thinking than likely principled outcome.

## References

David M. Blei, Andrew Y. Ng, and Michael I. Jordan. 2003. Latent Dirichlet allocation. J. Mach. Learn. Res. 3, null (3/1/2003), 993–1022.

Luke Breitfeller, Emily Ahn, & David Jurgens, & Yulia Tsvetkov,. (2019). Finding Microaggressions in the Wild: A Case for Locating Elusive Phenomena in Social Media Posts. 1664-1674. 10.18653/v1/D19-1176.

G. V. Cormack and M. R. Grossman. Evaluation of machine-learning protocols for technology-assisted review in electronic discovery. In SIGIR 2014.

Scott Deerwester, Susan T. Dumais, George W. Firmas, Thomas K. Landauer, and Richard Harshman. Indexing by latent semantic analysis. Journal of the American Society for Information Science, 41(6):391-407, September 1990.

Thomas Hofmann. 1999. Probabilistic latent semantic indexing. In Proceedings of the 22nd annual international ACM SIGIR conference on Research and development in information retrieval (SIGIR '99). Association for Computing Machinery, New York, NY, USA, 50–57. DOI:https://doi.org/10.1145/312624.312649

Herbert L. Roitblat, 2019. FOMO and eDiscovery. https://www.edrm.net/wp-content/uploads/2019/05/FOMO-and-eDiscovery-Herb-Roitblat-Ph.D..pdf